%% file: large_peaks.tex
\documentclass[a4paper,11pt]{article}
\pdfoutput=1

\usepackage{jcappub}

\usepackage[T1]{fontenc}

\usepackage{aas_macros}
\usepackage{slashed}

\title{\boldmath Local properties of the large-scale peaks of the CMB temperature}

\author[a,b]{A.~Marcos-Caballero,}
\author[a]{E.~Mart\'\i nez-Gonz\'alez}
\author[a]{and P.~Vielva}

\affiliation[a]{Instituto de F\'isica de Cantabria, CSIC-Universidad de Cantabria,\\ Avda. de los Castros s/n, 39005 Santander, Spain.}
\affiliation[b]{Dpto. de F\'isica Moderna, Universidad de Cantabria,\\ Avda. los Castros s/n, 39005 Santander, Spain.}

\emailAdd{marcos@ifca.unican.es}
\emailAdd{martinez@ifca.unican.es}
\emailAdd{vielva@ifca.unican.es}

\abstract{\input{abstract}}

\begin{document}
\maketitle
\flushbottom

\input{body}

\acknowledgments
Partial financial support from the Spanish Ministerio de Econom\'{i}a
y Competitividad Projects AYA2012-39475-C02-01 and Consolider-Ingenio
2010 CSD2010-00064 is acknowledged.

\bibliographystyle{JHEP}
\bibliography{large_peaks.bib}

\end{document}

%% file: abstract.tex
In the present work, we study the largest structures of the CMB
temperature measured by Planck in terms of the most prominent peaks on
the sky, which, in particular, are located in the southern galactic
hemisphere. Besides these large-scale features, the well-known Cold
Spot anomaly is included in the analysis. All these peaks would
contribute significantly to some of the CMB large-scale anomalies, as
the parity and hemispherical asymmetries, the dipole modulation, the
alignment between the quadrupole and the octopole, or in the case of
the Cold Spot, to the non-Gaussianity of the field. The analysis of
the peaks is performed by using their multipolar profiles, which
characterize the local shape of the peaks in terms of the discrete
Fourier transform of the azimuthal angle. In order to quantify the
local anisotropy of the peaks, the distribution of the phases of the
multipolar profiles is studied by using the Rayleigh random walk
methodology. Finally, a direct analysis of the 2-dimensional field
around the peaks is performed in order to take into account the effect
of the galactic mask. The results of the analysis conclude that, once
the peak amplitude and its first and second order derivatives at the
centre are conditioned, the rest of the field is compatible with the
standard model. In particular, it is observed that the Cold Spot
anomaly is caused by the large value of curvature at the centre.

%% file: body.tex
\section{Introduction}

The anisotropies in the Cosmic Microwave Background (CMB) are
described inside the standard model of cosmology, which assumes that
the initial perturbations are distributed according to a Gaussian in a
homogeneous and isotropic Universe. The recent measurements of the CMB
allow one to determine the cosmological parameters with high precision
\cite{planck132015}, showing an overall agreement between data and the
concordance cosmological model. Although, in general, this agreement
is large, some anomalies are found in the CMB at large scales. The
characterization of these deviations have an important role in
understanding the process which leads to the initial perturbations,
and hence, in the characterization of the inflationary model.

Different large-scale anomalies are found in the CMB, which were first
detected in the WMAP data (\cite{spergel2003}, see also references
below) and confirmed later by Planck \cite{planck162015}. Among them,
it is included the lack of correlations on large scales
\cite{bennett2003,copi2015b} which might leads to a low variance in
the temperature field \cite{monteserin2008}. In particular, in terms
of the angular power spectrum, the low variance anomaly can be seen as
a deficit of power in the lowest multipoles, in particular in the
quadrupole and octopole \cite{cruz2011}. Explanations in terms of an
early fast-roll phase of the inflation field preceding the standard
slow-roll phase have been proposed \cite{contaldi2003,destri2010}.

On the other hand, when the temperature field is expanded in terms of
the spherical harmonics, an unlikely alignment between the quadrupole
and the octopole is observed \cite{copi2015,planck232013}. The
interference between the two multipoles produce large-scale features
in the CMB temperature aligned with the ecliptic plane, which, in
particular, have more power in the southern hemisphere. The
hemispherical asymmetry, which is seen in the low multipoles range
($\ell \leq 64$), was first detected in \cite{eriksen2004} by looking
at the ratio of the power spectrum amplitude calculated on opposing
hemispheres, and later extended to include smaller scales in
\cite{hansen2009}. This particular asymmetry axis is also found in the
dipole modulation, which has been studied in real space
\cite{gordon2007,hoftuft2009}, as well as in harmonic space in order
to take into account the scale dependence in the analysis
\cite{planck162015}.

Moreover, parity asymmetries have been also analysed concluding that
there exists a difference in the variance of the odd and even
components of the temperature field
\cite{land2005,kim2010,gruppuso2011}. These studies have been extended
including a directional dependence in the parity asymmetry estimator
in several works \cite{naselsky2012,zhao2014}, indicating that the
preferred axis for parity violations could correspond to the
quadrupole and octopole alignment direction. Since the parity
asymmetry measures the correlation between antipodal points on the
sphere, this anomaly can be related to the lack of power at large
scales \cite{kim2011}.

In the present paper, we study the large-scale features on the CMB
temperature by identifying the most prominent peaks and analysing
their statistical properties. These largest peaks correspond to
structures located in the galactic southern hemisphere, more
precisely, in the quadrant where the south ecliptic pole is
located. This region of the sky corresponds to the direction where
some of the above mentioned anomalies are located (power asymmetry or
dipole modulation). Besides this directional asymmetries, the
interference of the quadrupole and the octopole induces an excess of
power in the ecliptic southern hemisphere which is caused by their
particular alignment \cite{copi2006}.  Additionally, although it is
not a peak as large as the others we consider, the Cold Spot
\cite{vielva2004,cruz2005} is also included in the analysis since it
presents an anomalous peak curvature. All these structures correspond
in part to the ``fingers'' and spots studied in
\cite{bennett2011}. Moreover, in a recent paper \cite{marcos2017}, a
multiscale analysis reveals that these peaks are the most outstanding
large-scale deviations in terms of either the amplitude or the
curvature.

The paper is organized as follows: in Section~\ref{sec:peak_theory},
the large-scale peaks are selected in the temperature field,
characterizing their local shape through the derivatives up to second
order. The analysis of the peaks is performed in terms of the radial
shape of the multipolar profiles in
Section~\ref{sec:multipolar_profiles}, whereas the study of their
phase correlations is considered in
Section~\ref{sec:phase_profiles}. In order to implement a partial sky
coverage properly, the work is completed with an analysis of the peaks
directly in real space. Finally, the summary and conclusions of the
paper are outlined in Section~\ref{sec:conclusions}.

\section{Characterization of the large-scale peaks}
\label{sec:peak_theory}

The peaks in the CMB correspond to local maxima or minima in the
temperature field, and they had been considered as useful geometrical
descriptors of the statistical properties of the primordial radiation
\cite{BBKS1986,bond1987,barreiro1997,komatsu2011,marcos2016}. In order
to have an extremum, constraints on the field derivatives have to be
imposed. Firstly, the critical point condition implies that the
gradient of the temperature must vanish at the peak location, but
additionally, in order to exclude possible saddle points, it is
imposed that the Hessian matrix is positive or negative definite,
depending whether the extremum is a minimum or a maximum. Therefore,
it is natural to characterize the peaks theoretically by conditioning
the first and second derivatives at the centre of the peak, as well as
the corresponding peak height. Following the notation in
\cite{marcos2016}, the derivatives on the sphere are calculated by
using the spin raising and lowering operators:
\begin{subequations}
\begin{equation}
\nu = \frac{T}{\sigma_\nu} \ ,
\end{equation}
\begin{equation}
\eta = \frac{\slashed{\partial}^*T}{\sigma_\eta} \ ,
\end{equation}
\begin{equation}
\kappa = -
\frac{\slashed{\partial}^*\slashed{\partial}T}{\sigma_\kappa} \ ,
\end{equation}
\begin{equation}
\epsilon = \frac{(\slashed{\partial}^*)^2T}{\sigma_\epsilon} \ ,
\end{equation}
\end{subequations}
where the derivatives are normalized in order to have dimensionless
quantities with unit variance. This set of parameters corresponds to
our peak degrees of freedom, which consist in two scalars ($\nu$ and
$\kappa$), one vector ($\eta$) and one $2$-spinor ($\epsilon$). Whilst
the value of the temperature at the extremum is given by the peak
height $\nu$, the local curvature is described by the Laplacian, which
is proportional to $\kappa$. On the other hand, the spinorial
quantities $\eta$ and $\epsilon$ represent the gradient and the
eccentricity tensor, respectively. The components of these two spinors
expressed in the helicity basis are given by complex numbers, whose
real and imaginary parts describe geometrical aspects of the peak. For
instance, the local eccentricity of the peak is proportional to the
modulus of $\epsilon$, whereas its phase represents the particular
direction of the principal axes on the sky. Regarding the gradient,
the real and imaginary parts of $\eta$ correspond to the components of
the first derivatives in the orthogonal local system of reference.
Theoretically, the gradient at the peak location must vanish by
definition, but we maintain this degree of freedom as non-zero in the
formalism because, in practice, there is a residual gradient due to
the fact that we are selecting peaks as local extrema in a discretised
field, which prevents us to impose $\eta = 0$. Although this non-null
value of the first derivative is very small compared to its standard
deviation, analysis based on conditioning the derivatives are very
sensitive to small variations of the conditioned values. In terms of
the peak shape, this effect causes a dipolar asymmetry at distances
larger than the peak size (see figure~\ref{fig:peaks_patches} in
Section~\ref{sec:real_space}).

\begin{figure}
\begin{center}
  \includegraphics[scale=0.70]{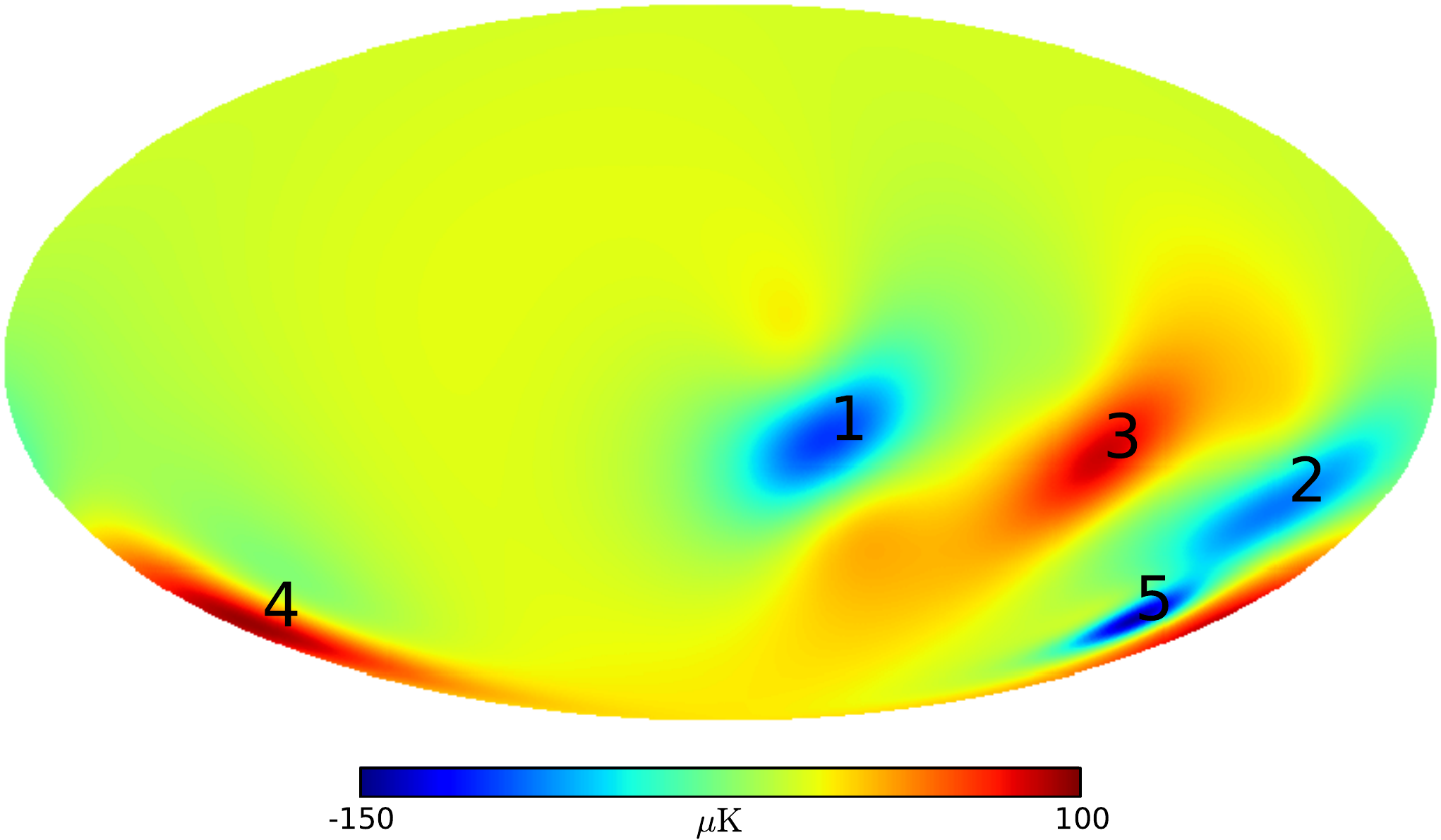}
\end{center}
\caption{Locations of the large-scale peaks considered in the paper,
  which are labeled with the numbers as referred in the text. The
  color map represents the theoretical mean field produced by
  conditioning the derivatives at the centre of the peaks. Notice that
  the correlation between peaks are not taken into account in this
  figure, causing that the derivatives do not correspond exactly to
  the observed values.}
\label{fig:peaks_map}
\end{figure}

In this work, we consider the five large-scale peaks given in
figure~\ref{fig:peaks_map}. In a recent paper \cite{marcos2017}, a
complete analysis of deviations on the derivatives fields is performed
at different scales, concluding that these peaks are the most
anomalous structures at large scales. The peaks labeled by $1$-$4$
correspond to two maxima and two minima selected in the CMB
temperature field filtered with a Gaussian with $R=10^\circ$.  These
peaks are the most prominent large-scale structures in the sky, which
are located in the same ecliptic hemisphere. The particular value of
$R = 10^\circ$ is chosen so that these large-scale fluctuations are
highlighted. On the other hand, the peak labeled by $5$ is the
well-known CMB anomaly called the Cold Spot \cite{cruz2005}, which is
included in the analysis because is the most outstanding large-scale
deviation in terms of the curvature \cite{marcos2017}, contributing to
the non-Gaussianity of the temperature field
\cite{vielva2004,cruz2005}. The Cold spot is usually characterized as
a minimum in the Spherical Mexican Hat Wavelet (SMHW)
\cite{martinez2002} coefficient map at the scale $R \approx
5^\circ$. As the SMHW is obtained from the Laplacian of the Gaussian
function, the coefficients map is equivalent to the curvature field
$\kappa$ of the temperature, filtered with a Gaussian with the same
scale than the SMHW. Since there is a strong correlation between $\nu$
and $\kappa$ at large scales, a minimum in the curvature corresponds
to a minimum in the temperature field itself. For this reason, we
equivalently define the Cold Spot as a minimum in $\nu$ at the scale
$R=5^\circ$.

In order to calculate the variances of the derivatives and other
theoretical quantities, a particular model has to be considered. The
following fiducial model is assumed: $\Omega_b h^2 = 0.2222$,
$\Omega_c h^2 = 0.1197$, $H_0 = 67.31 \ \mathrm{km/s}
\ \mathrm{Mpc^{-1}}$, $\tau = 0.078$, $n_s = 0.9655$ and $\ln(10^{10}
A_s) = 3.089$, which represent the Planck TT-lowP best-fit
cosmological parameters (\cite{planck132015}, table 3).

\begin{figure}
\begin{center}
  \includegraphics[scale=0.85]{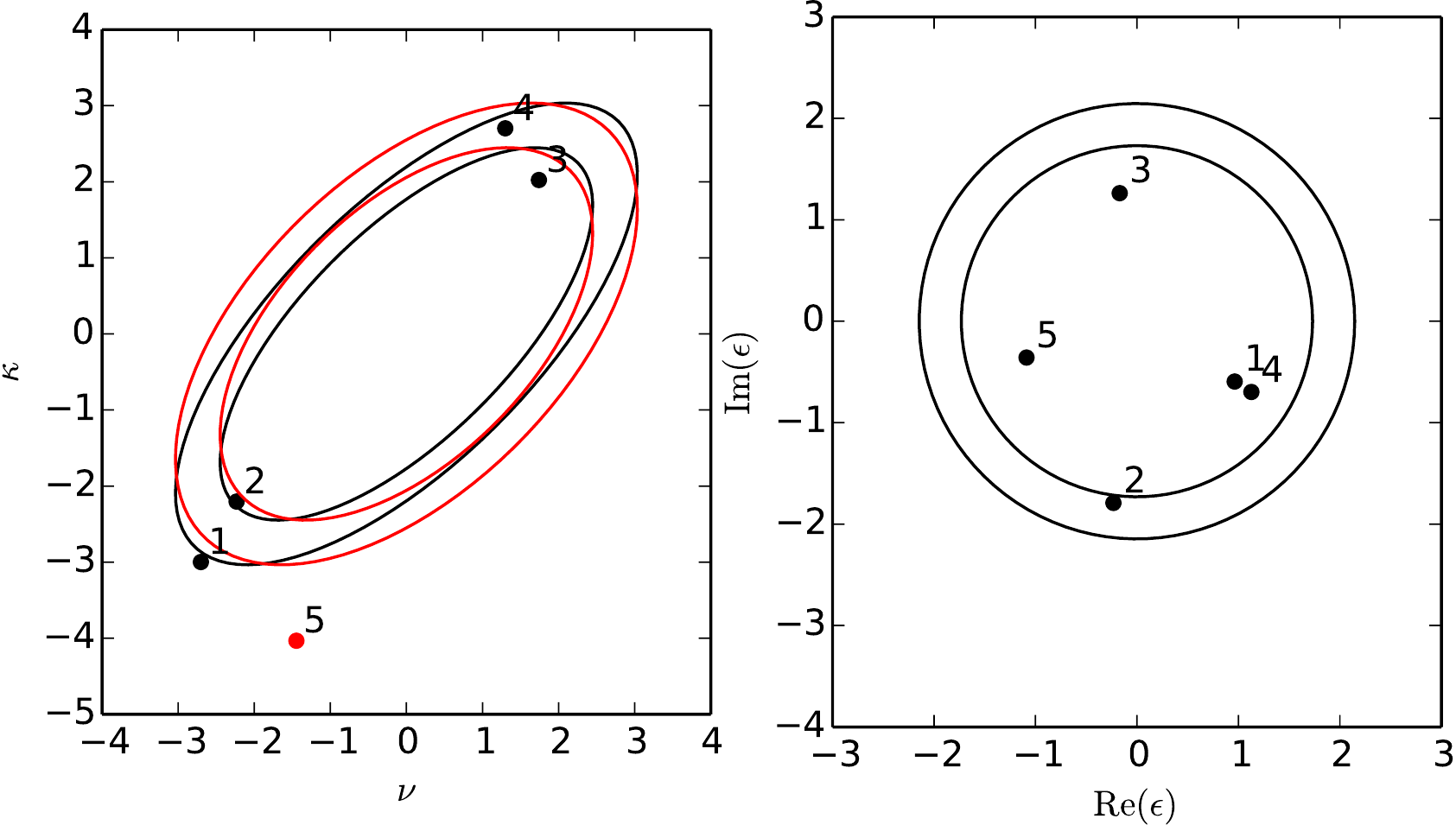}
\end{center}
\caption{The peak degrees of freedom labeled with the numbers which
  identify each peak throughout the paper. Whilst the peak height
  $\nu$ and the curvature $\kappa$ are shown in the left panel, the
  values of the eccentricity tensor $\epsilon$ are depicted in the
  complex plane in the right panel. The ellipses represent the
  probability contours at $95\%$ and $99\%$ levels. In the case of the
  $\nu$-$\kappa$ plane, the contours for the peaks $1$-$4$ are
  represented in black, and the corresponding ellipses for the Cold
  Spot (peak $5$) are shown in red.}
\label{fig:peak_dof}
\end{figure}

The values of the derivatives at the centre of the peaks obtained from
the Planck Commander map \cite{planck092015} are represented in
figure~\ref{fig:peak_dof}. The scalar degrees of freedom $\nu$ and
$\kappa$ are depicted in the same plane, showing the contours of the
one-point probability density function. Since the correlation between
$\nu$ and $\kappa$ depend on the scale where the peak is selected, it
is expected that the ellipses for the peaks $1$-$4$ ($R=10^\circ$) are
narrower than the ones for the peak $5$, whose scale is smaller
($R=5^\circ$). On the other hand, the one-point distribution of the
eccentricity tensor does not depend on the scale $R$
\cite{marcos2016}, and therefore the probability contours are the same
of all the peaks. The Cold Spot (peak $5$) is the peak which presents
a higher deviation in the $\nu$-$\kappa$ plane, mainly caused by the
large value of $\kappa \approx 4$. This value differs from the SMHW
coefficient $\kappa \approx 4.7$ reported in \cite{cruz2005} and
confirmed by \cite{planck162015} for the same scale, giving a lower
probability of finding a Cold Spot in the CMB temperature. The main
difference between these calculations is that, whereas in this work
the value of $\kappa$ is calculated by normalizing by the theoretical
variance $\sigma_{\kappa}$, in \cite{cruz2005} and \cite{planck162015}
the value of the SMHW coefficient is calculated by using the variance
estimated from the data, which is affected by the low variance of the
measured CMB field at large scales
\cite{monteserin2008,planck162015}. On the other hand, the
eccentricity of the Cold Spot is within the $2 \sigma$ level, which
implies that its shape is almost circular \cite{cruz2006}.

The deviation of the peaks derivatives with respect to the standard
model is considered by calculating the expected number of peaks with
$\nu$ and $\kappa$ as extreme as the corresponding observed values
which are present in one realization of the temperature field (see
\cite{marcos2016} for the expression of the number density of peaks on
the sphere). These numbers are $0.054$ for the Cold Spot and $0.14$
for the largest cold spot at $R=10^\circ$ (peak $1$), whereas the rest
of the peaks have an expected number per realization $\approx 1$. This
implies that a peak as extreme as the Cold Spot in terms of $\nu$ and
$\kappa$ is expected in every $19$ realizations of the CMB
temperature, given a more likely probability for the Cold Spot than
the calculation in \cite{cruz2005}, which considers a larger value for
the curvature $\kappa$ at the centre of the peak, as explained above.

\section{Multipolar profiles}
\label{sec:multipolar_profiles}                                   

In this section, we study the shape of the largest peaks observed in
the CMB temperature. Following the formalism of \cite{marcos2016}, the
shape of the peaks can be studied through the multipolar profiles,
which consist in the coefficients of the Fourier transform of the
azimuthal angle around the peak:
\begin{equation}
T_m(\theta) = \frac{1}{2\pi} \int \mathrm{d}\phi \ T(\theta,\phi)
e^{-im\phi} \ ,
\label{eqn:tm_theta}
\end{equation}
where the coordinates $\theta$ and $\phi$ represent the radial and
azimuthal coordinates, respectively, centered at the peak
location. The monopolar profile with $m=0$ corresponds to the standard
profile, which takes into account the spherical symmetric component of
the peak. On the other hand, the higher order profiles describe
different asymmetrical shapes, depending on the multipole $m$. For
instance, the profiles with $m=1$ and $m=2$ represent a dipole and a
quadrupole around the peak, respectively.

The derivatives at the centre of the peak affects to the local shape
depending on its spin. In particular, if the values of $\nu$,
$\kappa$, $\eta$ and $\epsilon$ are fixed at the centre, it is
obtained the following mean profiles \cite{marcos2016}:
\begin{subequations}
\begin{equation}
\langle T_0 (\theta) \rangle = \sum_{\ell=0}^\infty
\frac{2\ell+1}{4\pi} \left[ b_\nu + b_\kappa \ell (\ell+1) \right]
\ b_\ell w_\ell C_\ell \ P_\ell (\cos \theta) \ ,
\label{eqn:t0}
\end{equation}
\begin{equation}
\langle T_1 (\theta) \rangle = b_\eta \sum_{\ell=0}^\infty
\frac{2\ell+1}{4\pi} \ b_\ell w_\ell C_\ell \ P_\ell^1 (\cos \theta) \ ,
\end{equation}
\begin{equation}
\langle T_2 (\theta) \rangle = b_\epsilon \sum_{\ell=0}^\infty
\frac{2\ell+1}{4\pi} \ b_\ell w_\ell C_\ell \ P_\ell^2 (\cos \theta)
\ ,
\label{eqn:t2}
\end{equation}
\label{eqn:mean_profiles}
\end{subequations}
and $\langle T_m (\theta) \rangle = 0$ for $m \neq 0,1,2$. In these
equations, we have assumed that the peak is selected in the
temperature field filtered with the window function $w_\ell$, whereas
the profiles are calculated from a field observed with a beam
$b_\ell$. The bias parameters characterizing the mean profiles depend
on the particular values of the derivatives at the centre:
\begin{subequations}
\begin{equation}
b_\nu = \frac{\nu -\rho \kappa}{\sigma_\nu(1-\rho^2)} \ ,
\end{equation}
\begin{equation}
b_\kappa = \frac{\kappa -\rho \nu}{\sigma_\kappa(1-\rho^2)} \ ,
\end{equation}
\begin{equation}
b_\eta = \frac{\eta}{\sigma_\eta} \ ,
\end{equation}
\begin{equation}
b_\epsilon = \frac{\epsilon}{\sigma_\epsilon} \ ,
\end{equation}
\end{subequations}
where $\rho$ is the correlation coefficient between $\nu$ and
$\kappa$. As it is mentioned before, despite the fact that we are
selecting maxima and minima, the gradient at peak location does not
vanish because of the discretization of the field. This particular
residual affecting to the dipolar profile can be modelled as a small
bias $b_{\eta}$ depending on the measured value of $\eta$. This simple
modelization of the bias is enough to correct all the systematic
effect appearing in the subsequent analysis.

Additionally, when the local shape of the peak is fixed, the
covariance of the multipolar profiles are given by \cite{marcos2016}:
\begin{equation}
\langle T_m(\theta) T_{m^\prime}^*(\theta^\prime) \rangle = \langle T_m(\theta)
T_{m^\prime}^*(\theta^\prime) \rangle_{\mathrm{intr.}} + \langle T_m(\theta)
T_{m^\prime}^*(\theta^\prime) \rangle_{\mathrm{peak}} \ .
\label{eqn:tm_cov}
\end{equation}
Whilst the intrinsic part $\langle T_m(\theta) T_m(\theta^\prime)
\rangle_{\mathrm{intr.}}$ represents the covariance of the profile
when the derivatives at the centre are not constrained, the term
$\langle T_m(\theta) T_m(\theta^\prime) \rangle_{\mathrm{peak}}$ is
the modification of the covariance due to the fact of conditioning the
values of the derivatives. It is important to notice that the fact of
conditioning the derivatives to some particular values only affects to
the mean profiles, but not to the covariance. The intrinsic covariance
can be calculated from the angular power spectra in the following way
\cite{marcos2016}:
\begin{equation}
\langle T_m(\theta) T_{m^\prime}^*(\theta^\prime)
\rangle_{\mathrm{intr.}} = \delta_{mm^\prime} \sum_{\ell=m}^\infty
\frac{2\ell+1}{4\pi} \frac{(\ell-m)!}{(\ell+m)!} \ b_\ell^2 C_\ell
\ P_\ell^m(\cos\theta) P_\ell^m(\cos\theta^\prime) =
\delta_{mm^\prime} C_{m}^{\mathrm{intr.}}(\theta,\theta^\prime) \ .
\label{eqn:cov_intr}
\end{equation}
On the other hand, the contribution of the peak to the covariance of
the multipolar profiles is different form zero for $m=0,1,2$. In
general, it can be written as:
\begin{multline}
\langle T_m(\theta) T_{m^\prime}^*(\theta^\prime)
\rangle_{\mathrm{peak}} = \delta_{mm^\prime}
\sum_{\ell,\ell^\prime=m}^\infty \frac{2\ell+1}{4\pi}
\frac{2\ell^\prime+1}{4\pi} B_{\ell\ell^\prime}^m \ b_\ell w_\ell
C_\ell \ b_{\ell^\prime} w_{\ell^\prime} C_{\ell^\prime}
\ P_\ell^m(\cos\theta) P_{\ell^\prime}^m(\cos\theta^\prime) = \\ =
\delta_{mm^\prime} C_{m}^{\mathrm{peak}}(\theta,\theta^\prime) \ ,
\label{eqn:cov_peak}
\end{multline}
where the matrices $B_{\ell\ell^\prime}^m$ are given by:
\begin{subequations}
\begin{equation}
B_{\ell\ell^\prime}^0 = - \frac{1}{1-\rho^2} \left\{ \frac{1}{\sigma_\nu^2} -
\frac{\rho}{\sigma_\nu\sigma_\kappa}\left[\ell(\ell+1) +
  \ell^\prime(\ell^\prime+1)\right] +
\frac{\ell(\ell+1)\ell^\prime(\ell^\prime+1)}{\sigma_\kappa^2}\right\}
\ ,
\end{equation}
\begin{equation}
B_{\ell\ell^\prime}^1 = -\frac{1}{\sigma_\eta^2} \ ,
\end{equation}
\begin{equation}
B_{\ell\ell^\prime}^2 = -\frac{1}{\sigma_\epsilon^2} \ ,
\end{equation}
\end{subequations}
and $B_{\ell\ell^\prime}^m = 0$ for $m>2$. Since conditioning the
derivatives reduces the variance of the field, these coefficients are
always negative. These expressions can be generalised to consider
scenarios where only the amplitude or the curvature are
conditioned. In this case, we have that $B_{\ell\ell^\prime}^0$ equals
to $-\sigma_\nu^{-2}$ or $-\sigma_\kappa^{-2}$, depending on whether
$\nu$ or $\kappa$ is the conditioned variable.

As it is described in Section~\ref{sec:peak_theory}, the peaks $1$-$4$
are selected in a map filtered with a Gaussian with a scale $R =
10^\circ$, whereas the Cold Spot is defined as a peak in $R =
5^\circ$. Therefore, the window function $w_\ell$, which characterizes
the smoothing of the field where the amplitude and its derivatives are
calculated, is a Gaussian filter whose scale $R$ depends on the peak
considered. On the other hand, the filter $b_\ell$ corresponds to the
effective resolution of the maps over which the multipolar profiles
are calculated.

\begin{figure}
\begin{center}
  \includegraphics[scale=0.84]{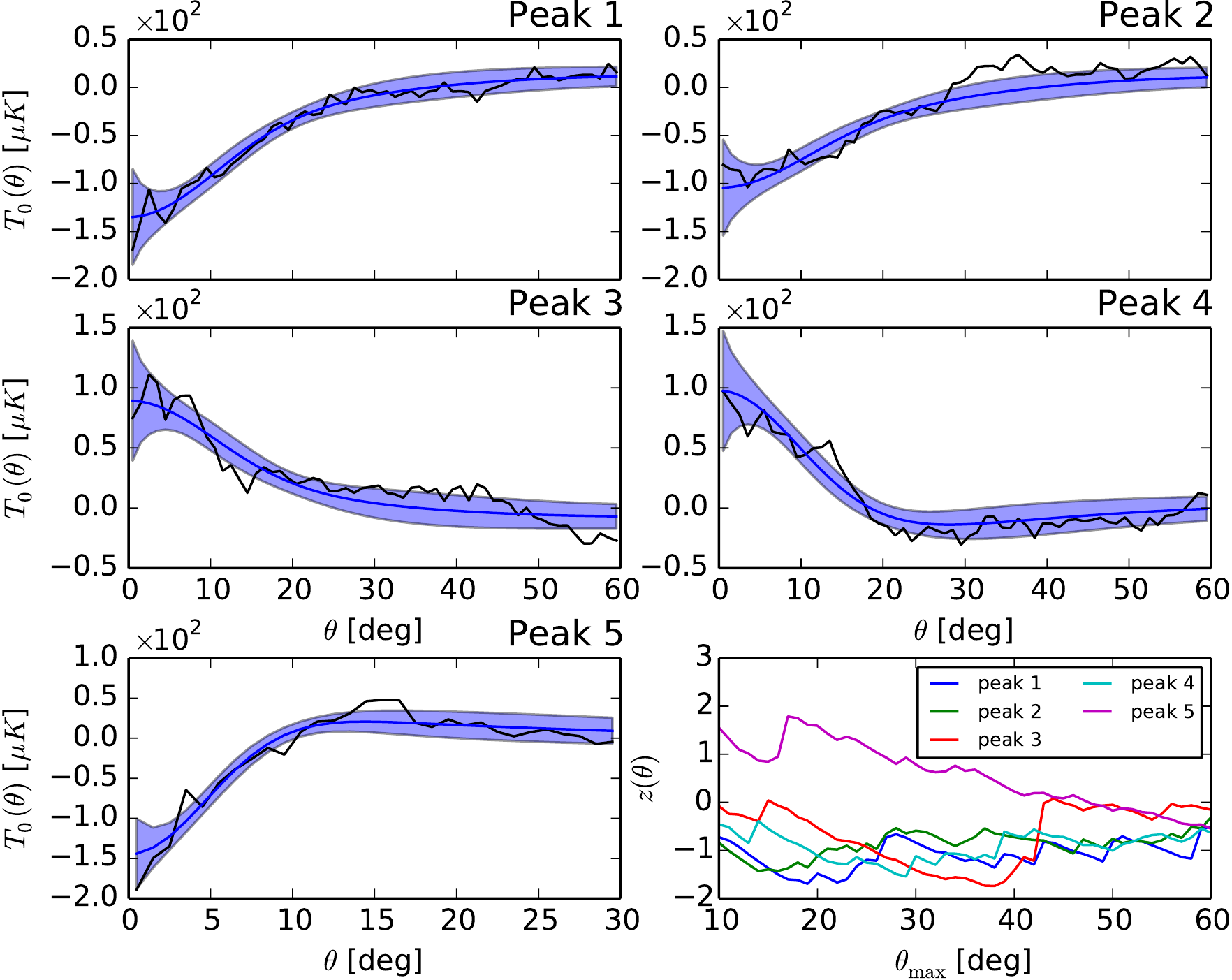}
\end{center}
\caption{Monopolar profiles ($m=0$) and their fit parameters $z$ (see
  the text for details) for the different peaks considered. The blue
  line represents the theoretical mean profiles conditioned to the
  values of $\nu$ and $\kappa$ at $\theta=0$ observed for each peak,
  and the shaded regions show the $1\sigma$ error bars. The fit
  parameter $z$ is depicted as a function of $\theta_{\mathrm{max}}$,
  the maximum value of $\theta$ of the profile considered in the fit.}
\label{fig:peaks_m0}
\end{figure}

\begin{figure}
\begin{center}
  \includegraphics[scale=0.84]{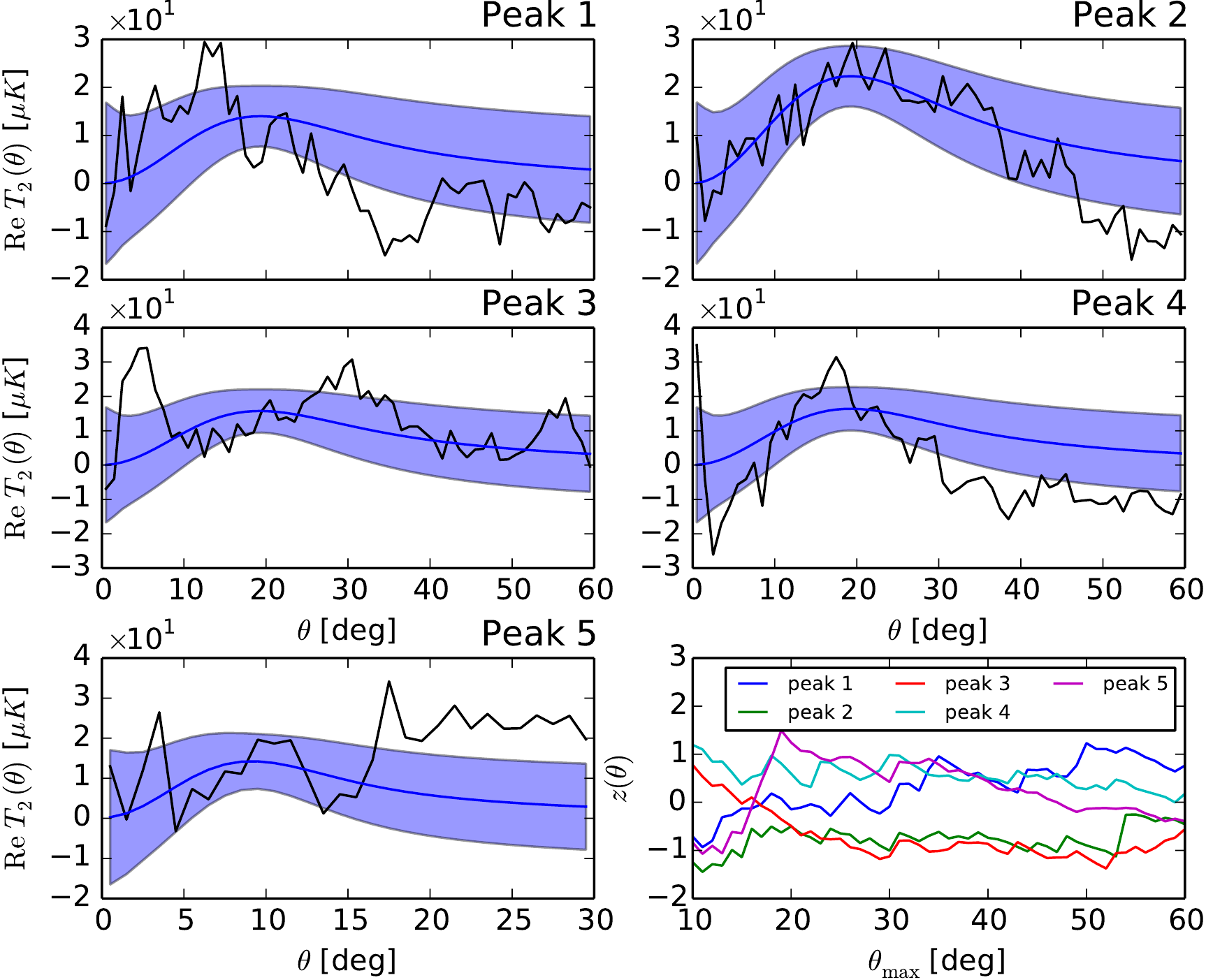}
\end{center}
\caption{Real part of the quadrupolar profiles ($m=2$) oriented along
  the principal axis for the different peaks considered. The blue line
  represents the theoretical mean profiles conditioned to the value of
  $\epsilon$ at $\theta=0$ observed for each peak, and the shaded
  regions show the $1\sigma$ error bars. Additionally, the fit
  parameter $z$ is depicted as a function of $\theta_{\mathrm{max}}$,
  the maximum value of $\theta$ considered in the fit.}
\label{fig:peaks_m2}
\end{figure}

In order to analyse the shape of the peaks, the values of $\nu$,
$\kappa$, $\eta$ and $\epsilon$ are conditioned to the measured values
at the centre of the peak. The observed monopolar, dipolar and
quadrupolar profiles are compared with the theoretical predictions in
figures~\ref{fig:peaks_m0}-\ref{fig:peaks_m2}. In the case of the
quadrupolar profile, a rotation is performed in order to align the
principal axes with the system of reference of the peak, such that
only the real part has non-zero expectation value. Statistical
deviations from the standard model are quantified using a $\chi^2$
test as a function of the maximum value of $\theta$ considered in the
analysis. Assuming that the CMB temperature is a Gaussian random
field, the conditional probability of the multipolar profiles obtained
fixing the values of the derivatives at the centre is also Gaussian,
therefore the $\chi^2$ test is appropriate for the analysis. It is
computed the following quantity for each $\chi^2$ value, which is
approximately normally distributed for a large number of degrees of
freedom:
\begin{equation}
z_m(\theta_{\mathrm{max}}) =
\frac{\chi^2_{m}(\theta_{\mathrm{max}})-n_f(\theta_{\mathrm{max}})}{\sqrt{2n_f(\theta_{\mathrm{max}})}}
\ ,
\end{equation}
where $\theta_{\mathrm{max}}$ represents the maximum value of $\theta$
considered in the test, and $n_f$ is the number of degrees of freedom
of the $\chi^2$ variable corresponding to that value of
$\theta_{\mathrm{max}}$. Whereas for $m=0$ the value of $n_f$ is equal
to the number of bins considered, for any other multipole $m$ it is
twice the number of bins due to the fact that the profiles take
complex values. In this equation, the statistics
$\chi^2_{m}(\theta_{\mathrm{max}})$ is computed from the measured
profiles and the theoretical mean profiles and covariances:
\begin{equation}
\chi^2_{m}(\theta_{\mathrm{max}}) = \left( 2 - \delta_{m0} \right)
\sum_{\theta_i,\theta_j \leq \theta_\mathrm{max}} \left[ T_m^*
  (\theta_i) - \langle T_m^*(\theta_i) \rangle \right]
C^{-1}_{m}(\theta_i,\theta_j) \left[ T_m(\theta_j) - \langle
  T_m(\theta_j) \rangle \right] \ ,
\end{equation}
where the matrix $C_{m}(\theta_i,\theta_j) =
C_{m}^{\mathrm{intr.}}(\theta_i,\theta_j) +
C_{m}^{\mathrm{peak}}(\theta_i,\theta_j)$ is the covariance between
the different bins of the multipolar profiles, which is given by the
sum of eq.~(\ref{eqn:cov_intr}) and eq.~(\ref{eqn:cov_peak}). The
summations in this expression are extended over the indices for which
the centre of the bins $\theta_i$ take values up to the
$\theta_\mathrm{max}$. Regarding the theoretical estimation, the mean
profiles and covariance must be also averaged in each bin in order to
compare with the data. This operation is equivalent to calculate the
integral of the associated Legendre functions in each interval of
$\theta$. In the literature \cite{didonato1982}, there exists
analytical formulae which allow to calculate these integrals
recursively (see appendix~\ref{app:plm_int}). Notice that both the
real and the imaginary parts of the multipolar profiles $T_m(\theta)$
are considered in the calculation of
$\chi^2_m(\theta_\mathrm{max})$. Commonly, the peaks are oriented
along the principal axes, in which case the mean value of the
imaginary parts vanishes.

Since we are interested in large-scale peaks, the galactic mask is a
problem in the calculation of the profiles, specially in the ones with
$m>0$, where the break of the isotropy of the field is
critical. Deconvolution techniques based on the Toeplitz matrix can be
used in order to correct the mask effect, but in the case of
aggressive masking the resulting profiles are not accurately
calculated. For this reason, we use an inpainted map without missing
pixels, more precisely, the CMB temperature field used in the analysis
of the multipolar profiles is the Planck Commander map
\cite{planck092015}, whose galactic mask region has been filled by
calculating a constrained Gaussian realization. On the other hand, in
Section~\ref{sec:real_space}, the peaks are analysed directly in real
space, where the missing pixels are not problematic, and therefore the
inpainting techniques are not required.

Following the expression in eq.~(\ref{eqn:tm_theta}), the multipolar
profiles are calculated by averaging the pixels in rings whose width
is $1^\circ$ and are centred at the different values of
$\theta$. Since the size of these bins is large compared with the
resolution of the Planck Commander map (FWHM $5'$), the contribution
of the instrumental noise can be ignored in our analysis. On the other
hand, the filter $b_\ell$ used in the calculation of the theoretical
profiles and covariance is the product of a Gaussian filter
characterizing the resolution of the data and the corresponding pixel
window function.

In figures~\ref{fig:peaks_m0}-\ref{fig:peaks_m2}, it is represented
the values of $z(\theta_{\mathrm{max}})$ for the monopolar, dipolar
and quadrupolar profiles, respectively. We can see that the deviation
of these profiles is less than $2\sigma$ in all the peaks
considered. Moreover, the multipolar profiles with values of $m$ up to
$10$ have been analysed, obtaining values which are compatible at
$3\sigma$ level with what is expected in the standard model.

\begin{figure}
\begin{center}
  \includegraphics[scale=0.84]{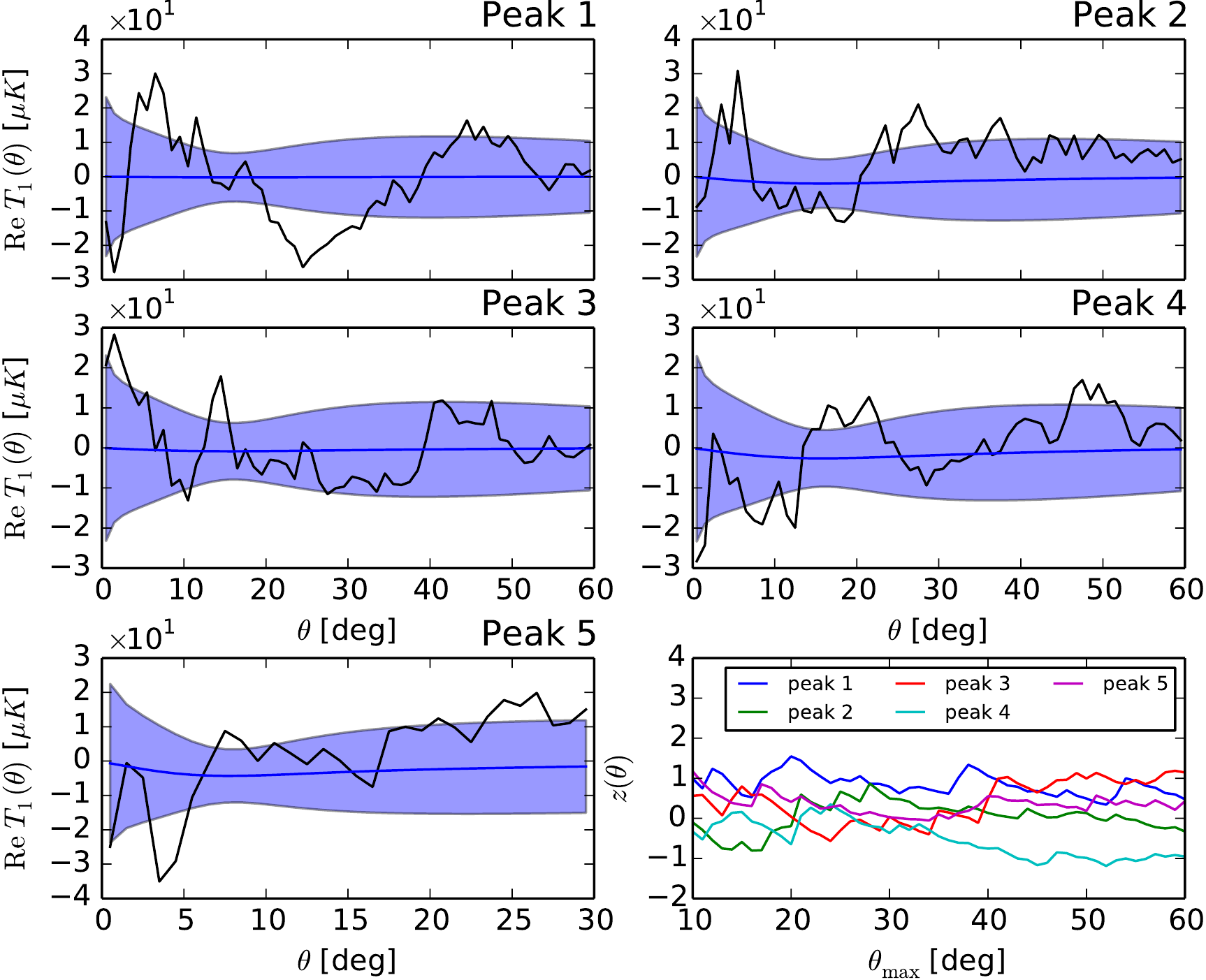}
\end{center}
\caption{Real part of the dipolar profiles ($m=1$) once the peaks are
  oriented in the direction of the residual gradient introduced by the
  pixelization. The blue line represents the theoretical mean profiles
  conditioned to the value of $\eta$ at $\theta=0$ observed for each
  peak, and the shaded regions show the $1\sigma$ error
  bars. Additionally, the fit parameter $z$ is depicted as a function
  of $\theta_{\mathrm{max}}$, the maximum value of $\theta$ considered
  in the fit.}
\end{figure}

\begin{figure}
\begin{center}
  \includegraphics[scale=0.40]{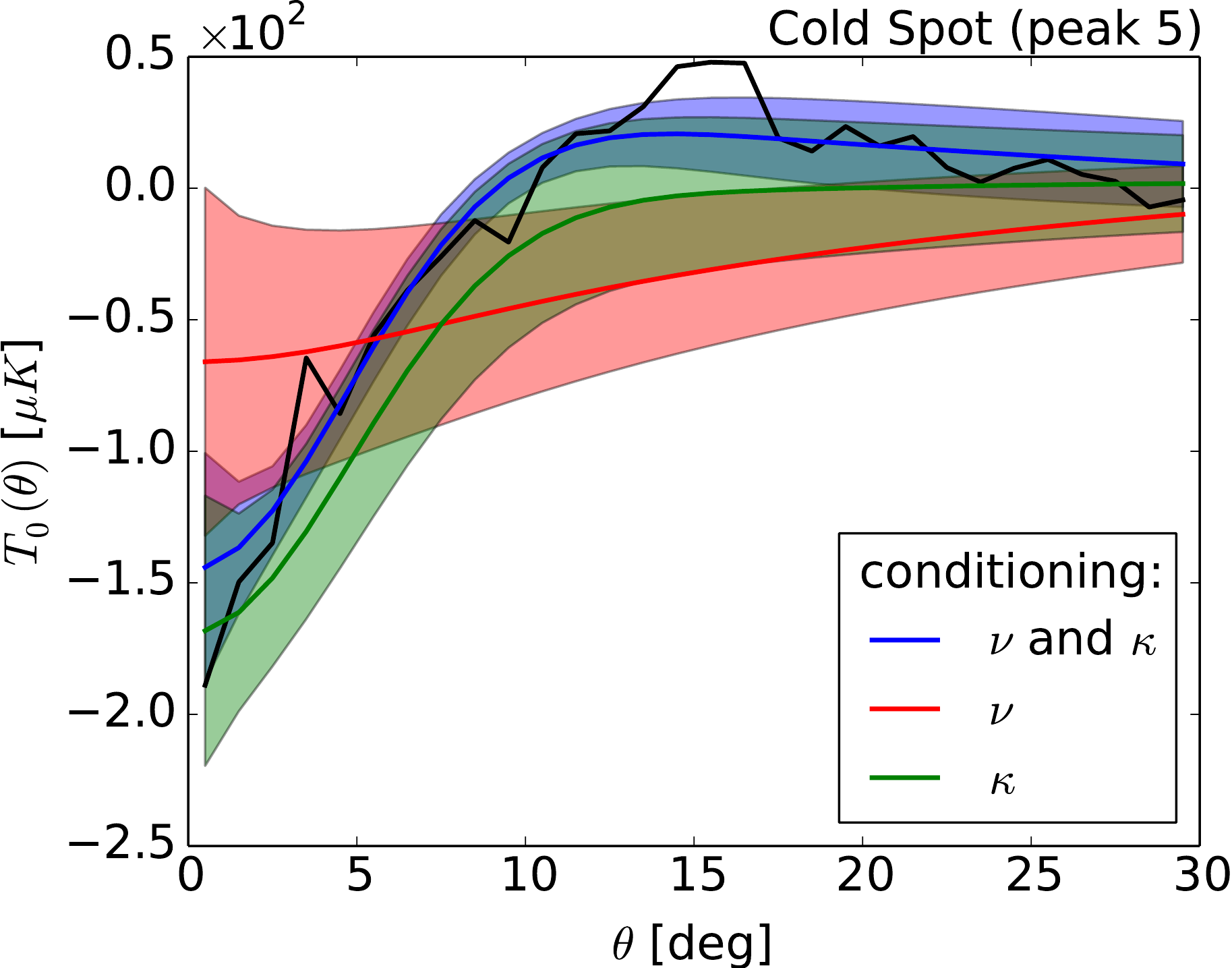}
  \includegraphics[scale=0.40]{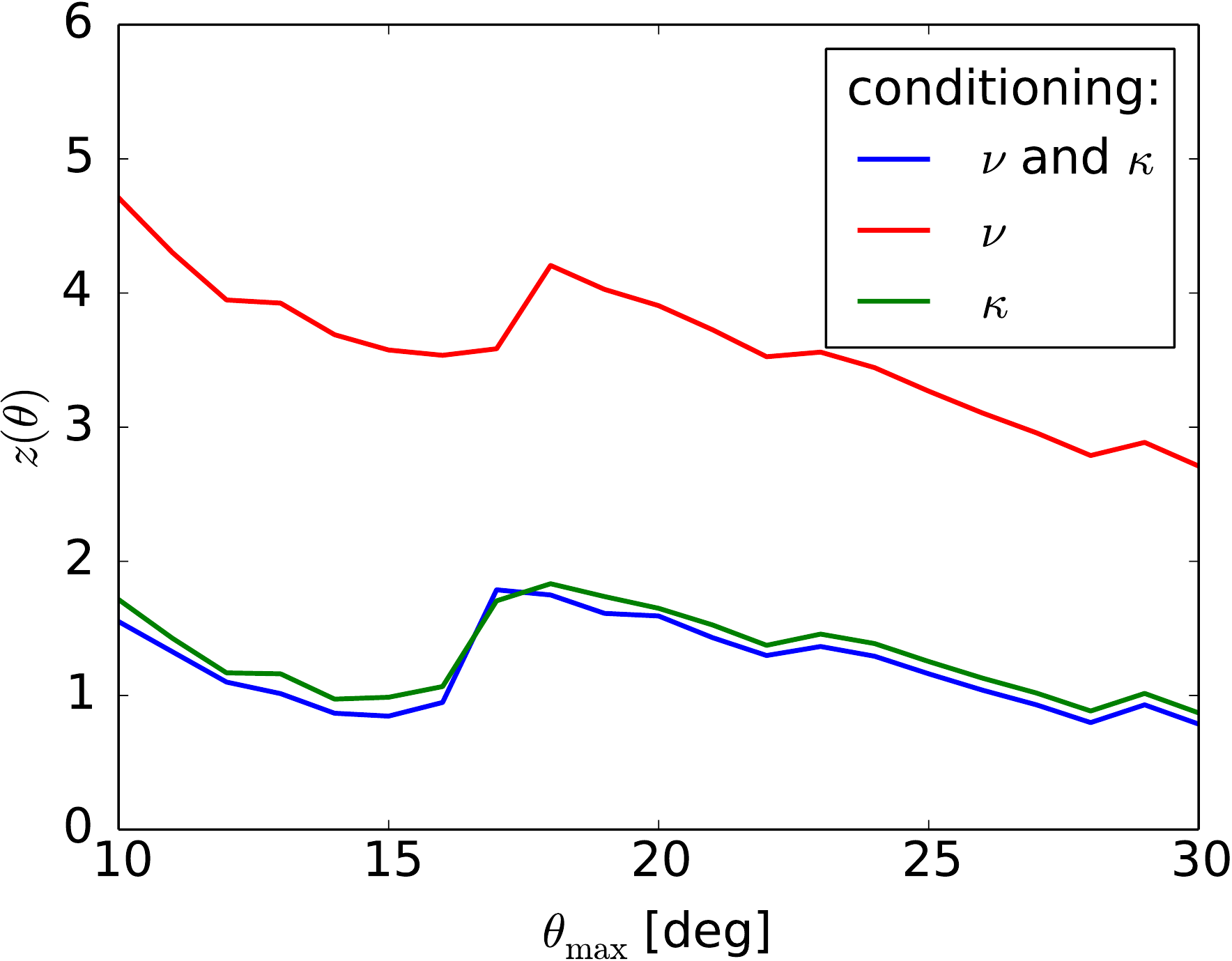}
\end{center}
\caption{\emph{Left:} Monopolar profiles of the Cold Spot obtained by
  conditioning the amplitude $\nu$, the curvature $\kappa$, or
  both. The shaded region represent the $1\sigma$ error bar in each
  case. \emph{Right:} $\chi^2$ test performed over these profiles
  measured in terms of the normal variable $z$ as a function of the
  maximum value of $\theta$ considered in the analysis.}
\label{fig:cs_profiles}
\end{figure}

Particularly, in the Cold Spot analysis, it is found that the
monopolar profile agrees with the standard model prediction when the
values of $\nu$ and $\kappa$ are conditioned. On the other hand, if
only the value of $\nu$ is fixed to the observed value whereas
$\kappa$ is averaged out using its probability density distribution
(see \cite{marcos2016}), the Cold Spot profile presents a $4.7\sigma$
deviation for $\theta < 10^\circ$ . This result implies that the Cold
Spot anomaly is mainly caused by the extremely large value of $\kappa$
at the centre, whereas when $\kappa$ is conditioned, no anomaly is
found in the monopolar profile. In figure~\ref{fig:cs_profiles}, it is
represented the monopolar profiles of the Cold Spot obtained by
conditioning, the peak height $\nu$, the curvature $\kappa$, or
both. Notice that the ring-shape in the Cold spot at $\approx
15^\circ$ is only recovered when both degrees of freedom are fixed,
which implies that this distinctive feature is produced by a combined
effect of a large value of $\kappa$ with a relatively small absolute
value of $\nu$.

\section{Phase correlations of the multipolar profiles}
\label{sec:phase_profiles}

In order to detect deviations from the standard model, the statistical
properties of the phases of the spherical harmonics coefficients have
been studied in several works. If the CMB temperature field is
non-Gaussian or anisotropic, correlations in the phases of the
$a_{\ell m}$'s may exists, which causes that they are not uniformly
distributed in the interval $[0,2\pi]$. There are different
statistical tests which can be applied to study the randomness of this
kind of periodic variables. For instance, the Kuiper's test, which is
a generalization of the KS test for circular data, has been used in
the analysis of the phases \cite{kovacs2013}. On the other hand, in
\cite{stannard2005}, the study of the Rayleigh statistics and the
random walk performed by the $a_{\ell m}$'s in the complex plane are
applied to the CMB temperature data. All the analyses considered in
these works are based on the spherical harmonics coefficients, which
describe the field in a particular system of reference, and therefore
their results could depend on the direction of the $z$ axis.
Additionally, in a previous work \cite{gott2007}, the genus of the
largest structures on the CMB ($\ell \leq 8$) are analysed concluding
that they corresponds to the ones derived from Gaussian field with
random phases. In the following, the phases of the multipolar
expansion centred at different peak locations are studied in terms of
the multipolar profiles.

The decomposition of the field around the peaks in terms of the
profiles $T_m(\theta)$ gives information about the contribution of
each multipolar pattern to the peak shape. In particular, the phases
of the multipolar profiles represent the orientation of each multipole
in the local system of reference centred at the peak. Given a
multipole $m$, the phases of $T_m(\theta)$ for different values of
$\theta$ are not independent due to the intrinsic correlations in the
field, and therefore an alignment of the multipoles is expected. In
order to test whether these correlations follow the standard model or
not, profiles whose bins in $\theta$ are independent are defined. More
precisely, considering $n$ bins of the radial angle labeled by
$\theta_a$, the following profiles are calculated:
\begin{equation}
\hat{T}_m(\theta_a) = \sum_{b=1}^a \lambda_{ab}^m \left[ T_m(\theta_b)
  - \langle T_m(\theta_b) \rangle \right]\ ,
\label{eqn:hat_tm}
\end{equation}
where the coefficients $\lambda_{ab}^m$ are chosen such that
$\hat{T}_m(\theta_a)$ have unit variance and no correlation for
different values of $a$. In practice, for each value of $m$, the
coefficients $\lambda_{ab}^m$ correspond to the components of the
lower triangular matrix obtained from the Cholesky decomposition of
the inverse covariance given in eq.~(\ref{eqn:tm_cov}). Additionally,
the mean profile $\langle T_m(\theta_b) \rangle$ is subtracted to the
data in eq.~(\ref{eqn:hat_tm}) in order to remove the peak degrees of
freedom from the phases analysis, since otherwise the phases can be
correlated because we are centred in a particular point of the field
with a peak. Notice that the cumulative sum in eq.~(\ref{eqn:hat_tm})
implies that $\hat{T}_m(\theta_a)$ only depends on the values of the
temperature with radial distance from the peak centre smaller that
$\theta_a$.

As the phases of $\hat{T}_m(\theta_a)$ are independent, they describe
a Rayleigh random walk in the complex plane for each value of $m$. At
the time step $N$, the position of this random walk is given by
\begin{equation}
Z_N^m = \sum_{a=1}^N \frac{\hat{T}_m(\theta_a)}{|\hat{T}_m(\theta_a)|}
\ .
\end{equation}
In these models of random walks, the time step $N$ corresponds to the
maximum radial angle $\theta_N$ considered in the multipolar
profile. Notice that, if a rotation of the system of reference around
the peak is performed with an angle $\alpha$, the positions of the
random walk transform as $Z_N^m e^{im\alpha}$, as can be deduced from
the transformation properties of the multipolar profiles. This is just
a rotation of angle $m\alpha$ of the complex plane where the random
walk moves on. Since the action of the rotation group on the steps
$Z_N^m$ is different for each value of $m$, we consider a random walk
for different multipolar profile separately. In previous works
\cite{stannard2005} based on the spherical harmonics coefficients,
different values of $m$ contribute to the steps, which implies that
the resulting random walk analysis is not invariant under rotations of
the $z$ axis. On the other hand, in the scenario considered in this
paper, the analysis of the random walks performed by the phases of
each multipolar profile only depends on the position of the peak, and
not on the orientation of the local system of reference.

The distance between the random walk position at the step $N$ and the
origin of the complex plane is approximately distributed following the
probability density
\begin{equation}
P_N(r) = \frac{2r}{N} \ e^{-r^2/N} \ ,
\label{eqn:prob_rw}
\end{equation}
which is valid for large values of $N$. From this equation, it can be
deduced that the variable $\sqrt{2/N}r$ is distributed according to
the Rayleigh distribution (or equivalently, $2r^2/N$ follows a
$\chi^2$ with two degrees of freedom). In order to achieve better
precision with this formula, the value of $r$ is calculated as follows
\cite{mardia2009}:
\begin{equation}
r_N^m = \sqrt{\left( 1 - \frac{1}{2N} \right) |Z_N^m|^2 +
  \frac{|Z_N^m|^4}{4N^2}} \ .
\end{equation}
For large values of $N$, the variable $r_N^m$ approach to the distance
travelled by the random walk $|Z_N^m|$. Considering this definition,
the variable $r_N^m$ follows the probability in
eq.~(\ref{eqn:prob_rw}) with $O(N^{-2})$ accuracy, instead of the
$O(N^{-1})$ error achieved with the standard definition of the
distance ($r_N^m = |Z_N^m|$).

\begin{figure}
\begin{center}
  \includegraphics[scale=0.84]{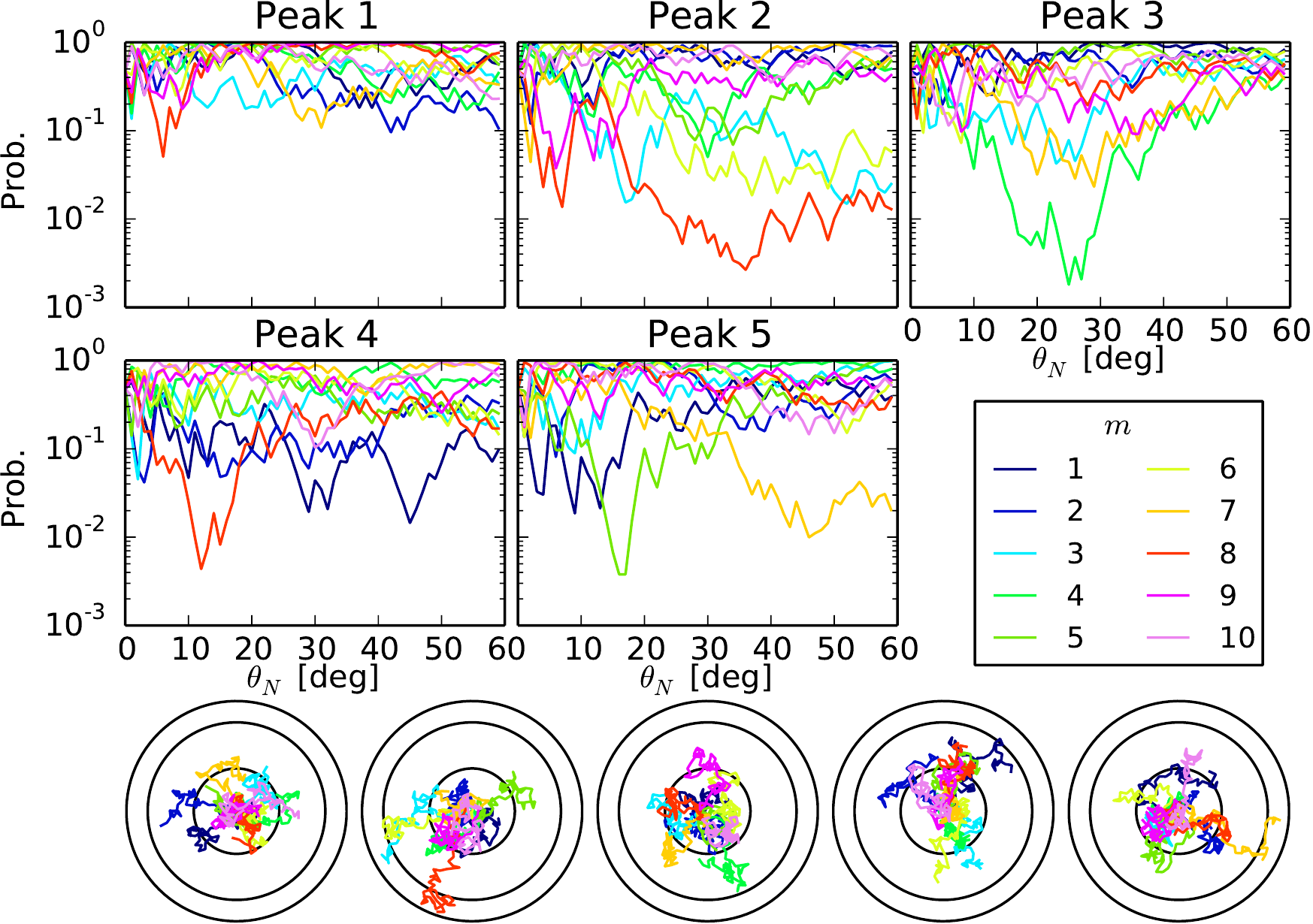}
\end{center}
\caption{{\emph{First and second rows:}} Lower tail probabilities of
  the distance travelled by the Rayleigh random walks $Z_N^m$ as a
  function of the time step $N$, which corresponds to the angle
  $\theta_N$ as labeled in the $x$ axis. Different colours represent
  the multipole $m$ of the profile as described in the
  legend. \emph{Bottom row:} Paths followed by the Rayleigh random
  walks derived by the phases of the multipolar profiles of the large
  scale peaks considered in the paper (numbers of the peaks are
  ordered from left to right). The black solid circles define the
  positions at which the probability distribution of the total
  travelled distance takes the values $0.50$, $0.95$ and $0.99$, from
  inner to the outer circle.}
\label{fig:phases_rw}
\end{figure}

The analysis is based on the fact that if the phases of the profiles
$\hat{T}_m(\theta_a)$ are correlated, the distances travelled by the
random walks will be greater than the ones expected from
eq.~(\ref{eqn:prob_rw}). The paths followed by the random walks
obtained from the phases of the multipolar profiles of the different
peaks considered are represented in figure~\ref{fig:phases_rw}. In
addition, the lower tail probability of the distance travelled by the
random walk at the time step $N$ is depicted.

It is possible to see some evidences of correlation of the phases for
the multipole $m=8$ in the peaks $2$ and $4$, whereas in the case of
the peaks $3$ and $5$, the most correlated multipoles are $m=4$ and
$m=5$, respectively. Regarding the Cold Spot (peak $5$), it is
important to notice that the maximum correlation is reached at
$\approx 15^\circ$, the angular distance which coincides with the
position of a hot ring around the centre of the Cold Spot. On the
other hand, the lack of anomalies in this analysis can be seen as
evidence of the low level of residuals in the Planck Commander map at
full-sky.

\section{Real space analysis}
\label{sec:real_space}

In the case of having an incomplete sky, the multipolar profiles with
$m \neq 0$ are very sensitive to the geometry of the mask. This is not
the case of the monopolar profiles ($m=0$), for which the standard sky
fraction correction is enough to have a good estimation of the
profile. Since we are interested in large-scale structures, it is very
unlikely to avoid the effect of the galactic mask in the analysis of
the peaks. In this section, we consider a real space approach,
analysing 2-dimensional patches around the peaks in a pixel-based
formalism. This allows one to take into account the mask in a simple
way, as compared with the Fourier analysis provided by the multipolar
profiles.

As in the previous sections, the patches around the peaks are
parametrized by the polar coordinates $(\theta,\phi)$ with the peak
located at the centre of the system of reference. If the peak is
described by its derivatives up to second order, the mean value of the
field is given as Fourier expansion in terms of the multipolar
profiles with $m=0,1,2$ \cite{marcos2016}:
\begin{equation}
\langle T(\theta,\phi) \rangle = \langle T_0(\theta) \rangle + \langle
T_1(\theta) \rangle e^{i\phi} + \langle T_1^*(\theta) \rangle
e^{-i\phi} + \langle T_2(\theta) \rangle e^{i2\phi} + \langle
T_2^*(\theta) \rangle e^{-i2\phi}
\end{equation}
On the other hand, the covariance of the temperature around the peak
can be decomposed in terms of the intrinsic covariance of the field,
and the covariance due to the effect of the peak selection
\cite{marcos2016}:
\begin{equation}
C(\theta,\phi,\theta^\prime,\phi^\prime) =
C_{\mathrm{intr.}}(\theta,\phi,\theta^\prime,\phi^\prime) +
C_0^\mathrm{peak}(\theta,\theta^\prime) + 2
C_1^\mathrm{peak}(\theta,\theta^\prime) \cos(\phi-\phi^\prime) + 2
C_2^\mathrm{peak}(\theta,\theta^\prime) \cos[2(\phi-\phi^\prime)] \ ,
\end{equation}
where $C_{\mathrm{intr.}}$ is the standard correlation function
between the points $(\theta,\phi)$ and $(\theta^\prime,\phi^\prime)$
of the temperature field, which does not consider the contribution of
the peak, and $C_m^\mathrm{peak}$ for $m=0,1,2$ (defined in
eq.~(\ref{eqn:cov_peak})) represents the contribution to the
covariance due to the fact that we are conditioning to the values of
the derivatives at the location of the peak.

\begin{figure}
\begin{center}
  \includegraphics[scale=0.70]{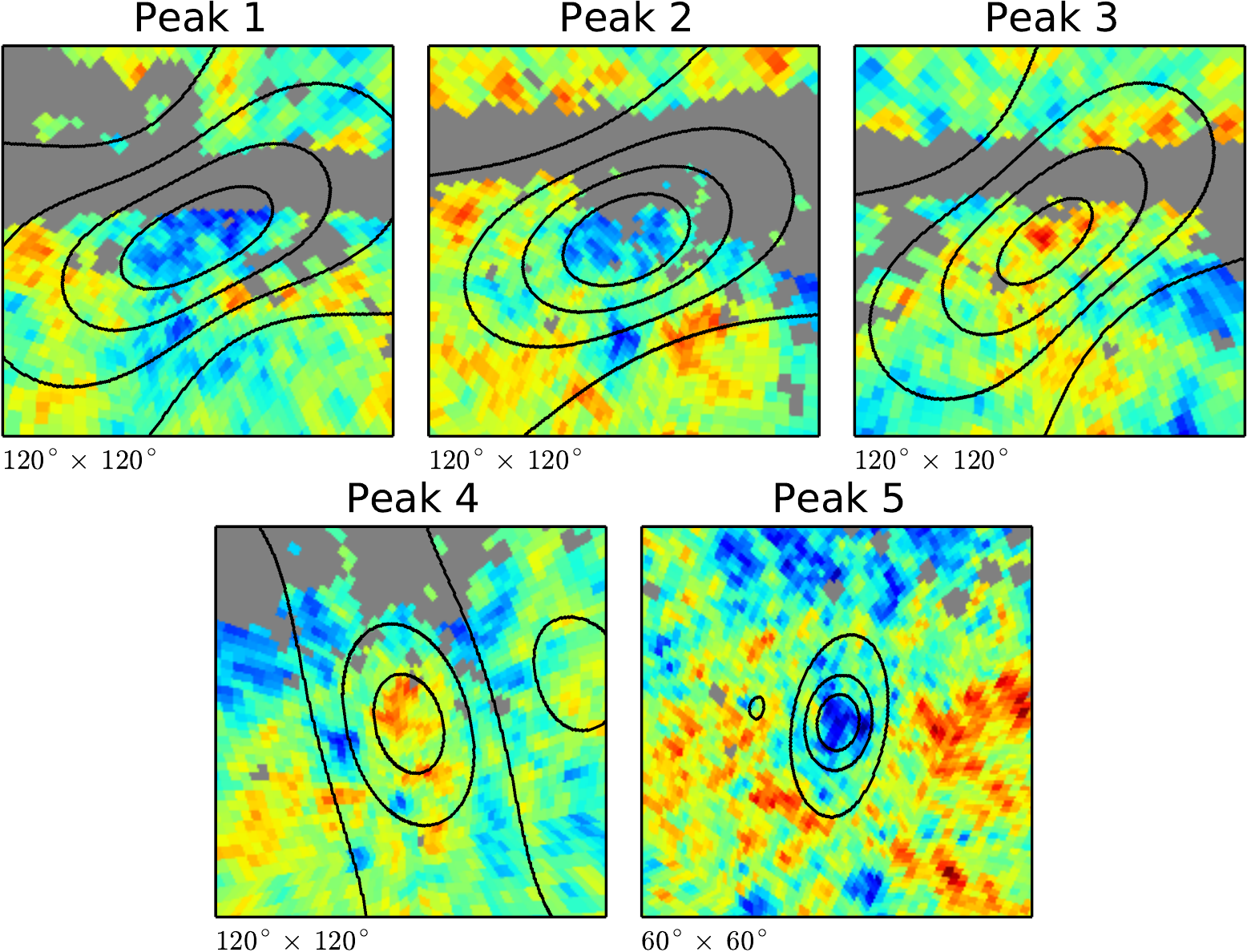}
  \includegraphics[scale=0.70]{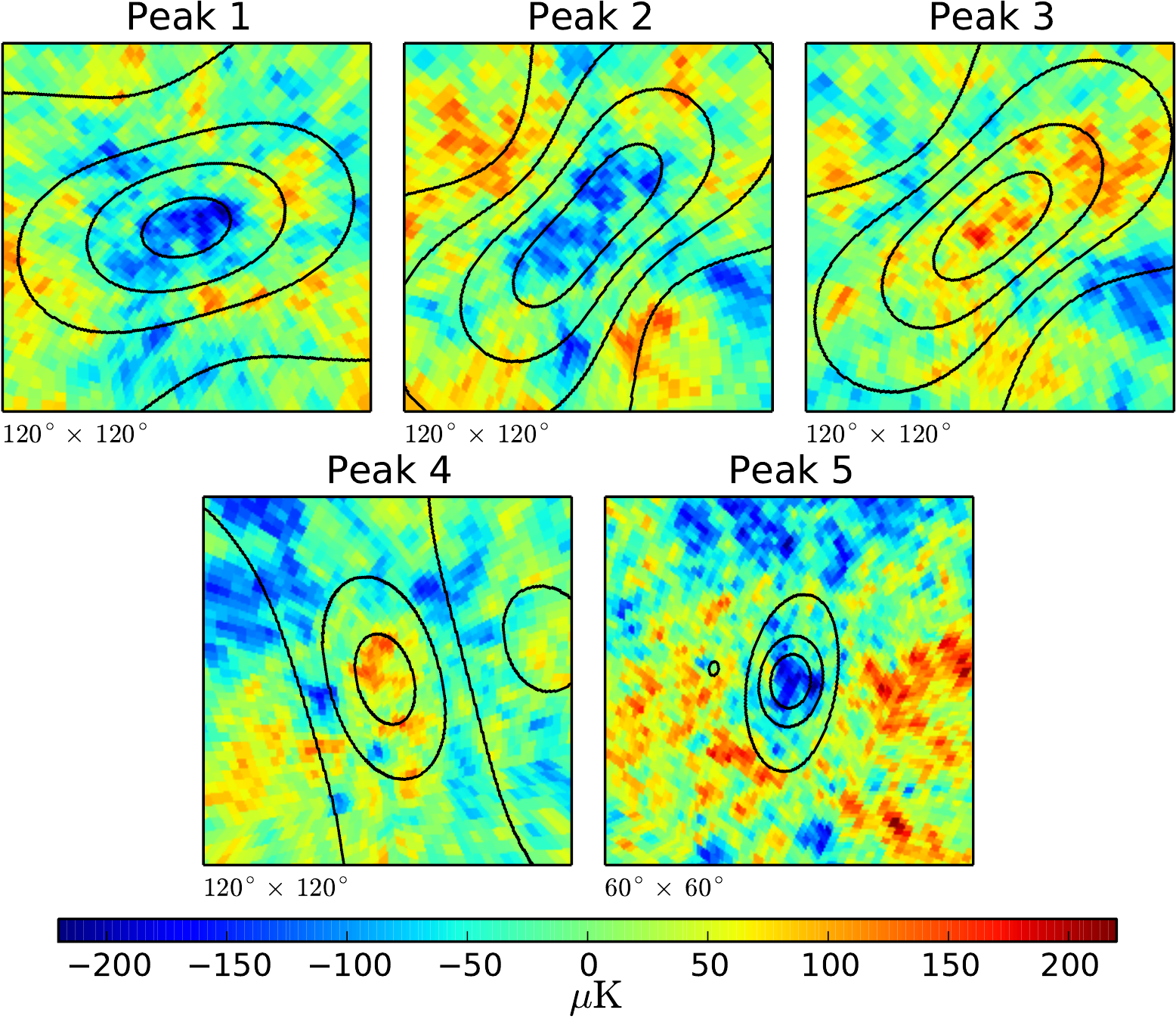}
\end{center}
\caption{Temperature patches of the Planck SEVEM (first and second
  rows) and Commander (third and fourth rows) maps around the peaks
  considered in the 2-dimensional analysis, where, in the case of
  SEVEM, the missing pixels due to the galactic mask are represented
  in gray. The contours depicts the theoretical mean temperature field
  obtained by conditioning the values of the derivatives at the centre
  of the peak.}
\label{fig:peaks_patches}
\end{figure}

The data used to study the peaks directly in the real space are the
Planck SEVEM and SMICA temperature maps, masked with their confidence
masks \cite{planck092015}. The analysis of the 2-dimensional patches
is based on the HEALPix pixelization scheme \cite{gorski2005} of the
regions around the peaks. For the largest peaks labelled by $1$-$4$,
the CMB data is filtered with a Gaussian of FWHM $2^\circ$ in harmonic
space and mapped at the resolution corresponding to $N_{\mathrm{side}}
= 32$. On the other hand, the Cold Spot (peak $5$) is analysed at
$N_{\mathrm{side}} = 64$ with a FWHM of $1^\circ$. The masks for the
two resolutions are calculated by smoothing the full resolution mask
with the corresponding Gaussian, and masking pixels below a given
threshold (in our case, $0.9$). The resulting patches consist in
disc-shape regions centred at the peaks with maximum radii of
$60^\circ$ for the largest peaks, and $30^\circ$ for the Cold Spot,
which leads to a total number of pixels $\sim 3000$ for each of the
peaks. The gnomic-projected CMB data at the peaks locations are
represented in figure~\ref{fig:peaks_patches}. Finally, the covariance
and the theoretical profiles are calculated by using
eqs.~(\ref{eqn:cov_peak}) and (\ref{eqn:mean_profiles}), where, as
in the case of the multipolar profiles, the window function $b_\ell$
is given by the map resolution considered and the pixel window
function.

The patches obtained from the data are compared with the theoretical
models of the peaks obtained after conditioning to the values of the
derivatives at the centre. The goodness of fit is evaluated by using a
$\chi^2$ test as a function of the maximum value of the radius
considered in the analysis. No significant deviations from the
theoretical models are found in the data for any of the peaks, a
result which is consistent with the analysis of the multipolar
profiles in Section~\ref{sec:multipolar_profiles}. Finally, in order
to check the consistency between the real space and multipolar profile
methodologies, the analysis of patches is repeated with the full-sky
Commander map, finding that both analysis are compatible with the
standard model.

\section{Conclusions}
\label{sec:conclusions}

In this work, we have studied the most prominent large-scale peaks in
the CMB temperature in terms of the multipolar profiles for different
values of $m$. Since the peaks are characterized by their derivatives
up to second order at the centre, we pay special attention to the
monopolar and quadrupolar profiles, which have expectation values
different form zero in this situation. Once the theoretical mean
profiles and covariances are calculated by conditioning the
derivatives to the observed values, a $\chi^2$ test is performed for
each peak and value of $m$. The analysis suggests that the theoretical
monopolar and quadrupolar profiles derived from the standard model
present a good agreement with the profiles obtained from the
data. Moreover, a broader analysis of the multipolar profiles
concludes that there is no significant deviations in the profiles with
$m$ up to $10$. These results implies that there is no anomalies in
the shape of the peaks considered, at least once the values of the
derivatives are conditioned.

The Cold Spot anomaly previously described in \cite{cruz2005} is
considered as a deviation in the Laplacian of the temperature field at
the smoothing scale $R=5^\circ$. The analysis performed by
conditioning both the peak height $\nu$ and the curvature $\kappa$
does not indicate any anomaly in the Cold Spot monopolar profile, but,
on the other hand, if only the value of $\nu$ is fixed, the profile
exhibits a $4.7 \sigma$ deviation up to a radius $\theta =
10^\circ$. This result shows that the Cold Spot anomaly is mainly
caused by the extremely large value of $\kappa$ at the centre, while
the field around it seems to be compatible with the Gaussian
correlations in the standard model. Moreover, it is observed that the
hot ring in the Cold Spot around $15^\circ$ is caused by a combination
of the large value of $\kappa$ with a comparatively small peak
amplitude $\nu$.

The study of the multipolar profiles is completed by analysing their
phases, which take into account the orientation of the different
multipolar shapes around the peaks. In general, even in the case of a
statistically isotropic field, the phase of the multipolar profiles
are correlated for different values of theta. For this reason, in the
paper, it is introduced an estimator which associates a
phase-independent profile $\hat{T}_m(\theta)$ to each multipolar
profile $T_m(\theta)$, given a fiducial model for its covariance. This
allows to define a Rayleigh random walk in terms of the phases of the
profiles, which moves as the value of $\theta$ increases. Statistical
deviations from the standard model are characterized by the total
length travelled by the random walk at a given time. If the distance
covered by the random walk associated to a given multipolar profile is
too large (too small), it means that the corresponding multipolar
profile of the peak has a correlation (anti-correlation) for different
values of $\theta $ which is greater than the one expected in the
standard model, and therefore the peak presents an alignment for that
value of $m$ not compatible with an isotropic field. Some alignments
are observed in few multipolar profiles of some of the large-scale
peaks considered. In particular, the Cold Spot presents an alignment
of the $m=5$ profiles which is maximum at the hot ring position
($15^\circ$).

Finally, the peaks are directly analysed in the real space by
considering $2$-dimensional patches around them. This methodology
allows to take into account a galactic mask, which cannot be done in
the multipolar profile expansion due to the spurious signal introduced
in that case. As in the profile analysis, the peak field is compared
with the theoretical expectation when the derivatives of the peak are
conditioned. In particular, the direction of the elongation of each
peak is fixed according to the observed eccentricity tensor. In this
case, the results are compatible with the ones obtained in the
multipolar profile analysis, concluding that the effect of the mask
does not change the main conclusions already found with the multipolar
profiles of the large-scale peaks.

\appendix

\section{Binning of the theoretical profiles}
\label{app:plm_int}

In order to compare with the data, the theoretical profiles have to be
binned in intervals of $\theta$. Since these profiles are expressed in
terms of the associated Legendre functions, this operation can be
done by calculating the following indefinite integrals:
\begin{equation}
I_\ell^m = \int \bar{P}_\ell^m(x) \ \mathrm{d}x =
\sqrt{\frac{2\ell+1}{4\pi}} \sqrt{\frac{(\ell-m)!}{(\ell+m)!}} \int
P_\ell^m (x) \ \mathrm{d}x \ ,
\end{equation}
where the normalised associated Legendre functions $\bar{P}_\ell^m$
are introduced in order to prevent from large numbers in the
calculations. On the one hand, the integrals for $m=0,2$ are
calculated from the Legendre polynomials \cite{didonato1982}:
\begin{subequations}
\begin{equation}
I_\ell^0(x) = \frac{1}{\sqrt{(2\ell+1)(2\ell+3)}} \bar{P}_{\ell+1}(x)
- \frac{1}{\sqrt{(2\ell+1)(2\ell-1)}} \bar{P}_{\ell-1}(x) \ ,
\end{equation}
\begin{equation}
I_\ell^2(x) = \frac{1}{\sqrt{(\ell+2)(\ell+1)\ell(\ell-1)}} \left[ -2
  I_\ell^0(x) + (\ell+3) x \bar{P}_\ell(x) - (\ell+1)
  \sqrt{\frac{2\ell+1}{2\ell+3}} \bar{P}_{\ell+1}(x) \right] \ .
\end{equation}
\end{subequations}
On the other hand, in the case of $m=1$, the integral can be
determined recursively following the expressions in
\cite{didonato1982}:
\begin{equation}
I_\ell^1(x) = \frac{\ell-2}{\ell+1}
\sqrt{\frac{(2\ell+1)\ell(\ell-2)}{(2\ell-3)(\ell+1)(\ell-1)}}
I_{\ell-2}^1(x) + \frac{1}{\ell+1}
\sqrt{\frac{(2\ell+1)(2\ell-1)}{(\ell+1)(\ell-1)}} \left( 1 - x^2
\right) \bar{P}_{\ell-1}^1(x) \ ,
\end{equation}
where the initial conditions are given by
\begin{subequations}
\begin{equation}
I_1^1(x) = - \frac{1}{4} \sqrt{\frac{3}{2\pi}} \left[ x \sqrt{1-x^2} +
  \arcsin x \right] \ ,
\end{equation}
\begin{equation}
I_2^1(x) = \frac{2}{3} \sqrt{1-x^2} \bar{P}_2^2(x) \ .
\end{equation}
\end{subequations}

Finally, the averaged value of the associated Legendre function
$P_\ell^m$ in the interval $[\theta_1,\theta_2]$ is expressed as
\begin{equation}
\frac{1}{\cos\theta_2-\cos\theta_1} \int_{\cos\theta_1}^{\cos\theta_2}
\mathrm{d}x \ P_\ell^m(x) = \sqrt{\frac{4\pi}{2\ell+1}}
\sqrt{\frac{(\ell+m)!}{(\ell-m)!}}
\frac{I_\ell^m(\cos\theta_2)-I_\ell^m(\cos\theta_1)}{\cos\theta_2-\cos\theta_1}
\ .
\end{equation}
Alternatively, these integrals can be evaluated using numerical
quadrature methods, but the recursive expressions above are faster and
more accurate.